\documentclass[english,tightenlines,onecolumn,showkeys,amsfonts,amsmath,amssymb,nofootinbib,a4paper,11pt]{revtex4-1}
\usepackage[english]{babel}
\usepackage{amsfonts}
\usepackage{amsmath}
\usepackage{amssymb}
\usepackage{graphicx}
\usepackage{enumerate}
\usepackage{hyperref}
\usepackage{color}

\makeatletter
\def\p@subsection{}
\makeatother

\def\Journal#1#2#3#4#5#6#7#8{#1: #3, #4. #2;\textbf{#5}:#6-#7. DOI: #8}
\def\JournalE#1#2#3#4#5#6#7{#1: #3, #4. #2;\textbf{#5}:#6. DOI: #7}

\newcommand{\RR}{\mathbb{R}}

\newcommand{\ZZ}{\mathbb{Z}}

\newcommand{\HH}{\mathcal{H}}

\begin{document}

\title{Invariance in Quantum Walks}
\author{Miquel Montero}
\affiliation{Department of Fundamental Physics, University of Barcelona (UB), Mart\'{\i} i Franqu\`es 1, E-08028 Barcelona, Spain\\ \tt{miquel.montero@ub.edu}}
\date{\today}


\begin{abstract}
In this Chapter, we present some interesting properties of quantum walks on the line. We concentrate our attention in the emergence of invariance and provide some insights into the ultimate origin of the observed behavior. In the first part of the Chapter, we review the building blocks of the quantum-mechanical version of the standard random walk in one dimension. The most distinctive difference between random and quantum walks is the replacement of the random coin in the former by the action of a unitary operator upon some internal property of the later. We provide explicit expressions for the solution to the problem when the most general form for the homogeneous unitary operator is considered, and we analyze several key features of the system as the presence of symmetries or stationary limits. After that, we analyze the consequences of letting the properties of the coin operator change from site to site, and from time step to time step. In spite of this lack of homogeneity, the probabilistic properties of the motion of the walker can remain unaltered if the coin variability is chosen adequately. Finally, we show how this invariance can be connected to the gauge  freedom of electromagnetism.
\end{abstract}
\maketitle

\section{I\lowercase{ntroduction}}

In their origins~\cite{ADZ93,TM02,NK03,JK03,VA12}, quantum walks (QWs) were thought as the quantum-mechanical generalization of the standard random walk in one dimension: the mathematical model describing the motion of a particle which follows a path that consists of a succession of jumps with fixed length whose direction depends on the random outcome of flipping a coin. In the quantum version, the coin toss is replaced by the action of a unitary operator upon some intrinsic degree of freedom of the system, a quantum observable with only two possible eigenvalues: e.g., the spin of an electron, the polarization of a photon, or the chirality of a molecule.

After this preliminary analysis, it became clear that the similitude between these two processes was mainly formal, and that random and quantum walks displayed divergent properties~\cite{CFG03}. The most remarkable of these discrepancies is perhaps the ability of unbiased QWs to spread over the line, not as the square root of the elapsed time, the fingerprint of any diffusion process, but with constant speed~\cite{ABNVW01}. This higher rate of percolation enables the formulation of quantum algorithms~\cite{PS97,FG98} that can tackle some problems in a more efficient way than their classical analogues: For instance, QWs are very promising resources for optimal searching~\cite{SKW03,AMB10,MNRS11}. Today, QWs have exceeded the boundaries of quantum computation and attracted the attention of researchers from other fields as, for example, information theory or game theory~\cite{FAJ04,BFT08,CB11,RH11}.

As a consequence of this wide interest, diverse extensions of the discrete-time QW on the line have been considered in the past. Most of these variations are related with the properties of the unitary coin operator~\cite{CSL08}, backbone of the novel features of the process. Thus, one can find in the literature QWs whose evolution depends on more than one coin~\cite{BCA03a,TFMK03,VBBB05}, QWs that suffer from decoherence~\cite{BCA03b,KT03}, or QWs driven by inhomogeneous, site-dependent coins~\cite{WLKGB04,RASAD05,SK10,KLS13,ZXT14,XQTS14}. There are also precedents where the temporal variability of the QW is explicit: in the form of a recursive rule for the coin selection, as in the so-called Fibonacci QWs~\cite{RMM04,AR09a}, through a given function that determines the value of the coin parameters~\cite{AR09b,RS14,BNPRS06}, or by means of an auxiliary random process that modifies properties of the coin~\cite{AVWW11}.

The main goal in most of these seminal papers is to find out new and exciting features that the considered modifications introduce in the behavior of the system, like the emergence of quasiperiodic patterns or the induction of dynamic localization. Recent works~\cite{MBD13,MBD14,MM14}, however, have also regarded the issue from the opposite point of view, by exploring the conditions under which the evolution of the system results unchanged. In particular, Ref.~\cite{MM14} considers the case of a discrete-time QW on the line with a time-dependent coin, a unitary operator with changing phase factors.

These phase factors are three parameters that appear in the definition of the coin operator whose relevance has been sometimes ignored in the past: When these phases are static magnitudes, they are superfluous~\cite{MM15}, but if they are dynamic quantities, they can substantially modify the evolution of the system. This fact does not close the door to the possibility that a set of well-tuned variable phase factors can keep the process unchanged from a probabilistic perspective. This defines a control mechanism that can compensate externally-induced decoherence and introduces a nontrivial invariance to be added to other well-known symmetries of QWs~\cite{CSB07,JKA12,TK12}.

In this Chapter we will review the approach taken in~\cite{MM14} and consider a generalization of it. Now, the evolution of the discrete-time quantum walker on the line will be subjected to the introduction of a fully inhomogeneous coin operator: The properties of the unitary operator will depend both on the location and on the present time through the action of the aforementioned phase factors. This extra variability leads to additional constraints to be satisfied by these magnitudes if one wants to guarantee that the properties of the motion of the walker remain unaltered. Finally, we  will connect our results with those appearing in Ref.~\cite{MBD14}, where the authors considered how the inclusion of time- and site-dependent phase factors in the coin operator of a quantum walk on the line may induce some  dynamics which, in the continuous limit, can be linked with the propagation of a Dirac spinor coupled to some external electromagnetic field. We will also explore the implications of this mapping here.    

\section{F\lowercase{undamentals of} QW\lowercase{s}}
\label{Sec_Process}
We begin this Chapter with a survey of the fundamental concepts required in the designing of discrete quantum walks on the line. In its simplest version, the particle represented by the walker can occupy detached and numerable locations on a one-dimensional space. This space of positions may be just a topological space (a graph or a chain, for instance) or can be endowed with a metric. In such a case, it is usual to consider that the sites are separated by a fixed distance $\ell$, so that $X=n\cdot \ell$. Within this standard framework, time increases in discrete steps as well,~\footnote{There is another kind of quantum walk, called {\it continuous\/} quantum walk, in which the walker can modify its position at any time: this is the quantum counterpart of continuous-time random walk.  The evolution of processes belonging to this category is ruled by a Hamiltonian and the corresponding Schr\"odinger equation. In spite they are different, discrete and continuous quantum walks share common traits~\cite{NK05}. } $T=t\cdot\tau$, $\tau$ being the sojourn time so that variable $t$ becomes a non-negative integer index, $t\in\{0,1,2,\cdots\}$, and the evolution of the system is just a sequence of states, $|\psi\rangle_{t}$. 

Up to this point, there is no significant difference between random and quantum walks. The major distinction is found in the nature of the random event that determines the progress of the particle. While in a world governed by the laws of classical mechanics, randomness is the way in which we describe the uncertain effect of multiple (and usually uncontrollable) external agents acting upon a system, in the realms of quantum mechanics randomness is not an exogenous ingredient. This means that we can use some internal degree of freedom in the quantum system with two possible eigenvalues (the spin, the polarization or the chirality) as a proxy for the coin, and understand that any change in this inner property is the result of the act of tossing. Therefore, to represent the state of the walker we need two different Hilbert spaces: $\HH_P$, the Hilbert space of particle positions spanned by the basis $\left\{| n\rangle : n \in \ZZ\right\}$, and the Hilbert space of the coin states, $\HH_C$, which is spanned by the basis $\left\{|+\rangle, |-\rangle\right\}$. The expression of $|\psi\rangle_{t}$ in the resulting Hilbert space $\HH$, $\HH\equiv\HH_C\otimes \HH_P$, reads 
\begin{equation}
|\psi\rangle_{t} =\sum_{n=-\infty}^{\infty}\left[\psi^{ }_{+}(n,t) |+\rangle \otimes | n\rangle+ \psi^{ }_{-}(n,t)   |-\rangle\otimes | n\rangle \right],
\label{psi_t_gen}
\end{equation}
where we have introduced the wave-function components $\psi_{\pm}(n,t)$, the two-dimensional projection of the state of the walker into the elements of the basis:
\begin{eqnarray}
\psi^{ }_{+}(n,t)&\equiv& \langle n|  \otimes  \langle +| \psi\rangle_t, \label{Def_Psi_P}\\
\psi^{ }_{-}(n,t)&\equiv& \langle n|  \otimes  \langle -| \psi\rangle_t. \label{Def_Psi_M} 
\end{eqnarray}

Now we have to consider the mechanism that connects these two properties, position and quirality, which eventually leads to a model for the dynamics of $\psi_{\pm}(n,t)$. Evolution in the discrete-time, discrete-space quantum walk can be regarded as the result of the action of operator $\widehat{\mathcal{T}}$, $\widehat{\mathcal{T}}\equiv \widehat{\mathcal{S}}\, \widehat{\mathcal{U}}$, on the state of the system $|\psi\rangle_{t}$. As it can be observed, the practical implementation of operator $\widehat{\mathcal{T}}$ has two stages: In the first one, the unitary operator $\widehat{\mathcal{U}}$ modifies exclusively the internal degree of freedom of the quantum system, in what represents the throw of the coin as indicated above,
\begin{eqnarray}
\widehat{\mathcal{U}}
&\equiv& \sum_{n=-\infty}^{\infty}e^{i\chi}\big[e^{i \alpha} \cos \theta |+\rangle  \langle +| +e^{-i \beta} \sin \theta |+\rangle  \langle -| \nonumber \\
&+& e^{i \beta} \sin \theta |-\rangle  \langle +| - e^{-i \alpha} \cos \theta  |-\rangle  \langle -|\big]\otimes |n\rangle  \langle n|.%\nonumber \\
\label{U_coin_bas}
\end{eqnarray}
In a second step, the shift operator $\widehat{\mathcal{S}}$ {\it moves\/} the walker depending on the result obtained after the last toss:~\footnote{With the present definition, the problem is spatially homogeneous and the system displays translational invariance. Therefore, alternative shift rules may be considered with equivalent results, as in the case of directed quantum walks~\cite{HM09,MM13}, where the particle can either remain still in the place or proceed in a fixed direction but never move backward.}
\begin{equation}
\widehat{\mathcal{S}} \left(|\pm\rangle\otimes| n\rangle \right)= |\pm\rangle\otimes| {n\pm 1}\rangle.
\label{shift_operator}
\end{equation}
Therefore, the state of the system at a later time $|\psi\rangle_{t+1} $ is recovered by application of $\widehat{\mathcal{T}}$ to the preset state:   
\begin{equation}
|\psi\rangle_{t+1} =\widehat{\mathcal{T}}\,|\psi\rangle_{t},
\label{evol_t_bas}
\end{equation}
and the complete evolution of the system is determined once $|\psi\rangle_{0}\equiv|\psi\rangle_{t=0}$ is selected. As in any quantum problem, one can consider for the initial state of the walker any  combination of the elements in the basis of $\HH$, a configuration that may lead to some degree of uncertainty in the position and/or the chirality of the system. However, the interest in establishing parallelisms between classical and quantum walkers encourages the choice in which, at the beginning, the particle position is known exactly, but its internal degree of freedom is {\it aligned\/} arbitrarily:
\begin{equation}
|\psi\rangle_{0} =\left(\cos \eta |+\rangle + e^{i \gamma}\sin \eta   |-\rangle\right) \otimes | 0\rangle.
\label{psi_zero_gen}
\end{equation}
Needless to say that the linearity and the translational invariance of the problem ensure that the solution for a general initial state can be recovered by direct superposition of the evolution of Eq.~\eqref{psi_zero_gen}, see Eqs.~\eqref{Sol_Psi_P} to~\eqref{omega_r} below.

The similarities and dissimilarities between classical and quantum walks must be grounded on the analysis of the probability mass function (PMF) of the process, $\rho(n,t)$, the probability that the walker can be found in a particular position $n$ at a given time $t$. The PMF for a random walk is
\begin{equation}
\rho_{\rm clas.}(n,t)=\binom{t}{\frac{t+n}{2}}\,p^{\frac{t+n}{2}} (1-p)^{\frac{t-n}{2}},
\end{equation}
where $p$ is the probability of obtaining a {\it head\/} as the result of flipping the coin. For the quantum walk, $\rho(n,t)$ is the sum of the squared modulus of the wave-function components, 
\begin{equation}
\rho(n,t)=\left|\psi_{+}(n,t)\right|^2+\left|\psi_{-}(n,t)\right|^2.
\end{equation}
On the basis of the values of the moduli of $\psi_{\pm}(n,t)$ we can also express the probability of obtaining a {\it head\/} value or a {\it tail\/} value when measuring the global coin state of the walker:
\begin{equation}
P_{\pm}(t)\equiv \sum_{n=-\infty}^{\infty} \left|\psi_{\pm}(n,t)\right|^2,
\end{equation}
or the value of $M(n,t)$,
\begin{equation}
M(n,t) \equiv\left|\psi_{+}(n,t)\right|^2-\left|\psi_{-}(n,t)\right|^2,
\end{equation}
another interesting magnitude that can be connected with the local magnetization of the system %in the $z$ direction, 
if the internal degree of freedom has its origin in the spin of the particle~\cite{SA13}.

\subsection{General solution}

The evolution operator $\widehat{\mathcal{T}}$ induces the following set of recursive equations in the wave-function components,
\begin{eqnarray}
\psi^{ }_{+}(n,t)&=&e^{i\chi^{ }_{}} \big[e^{i\alpha}\cos \theta \,\psi_{+}(n-1,t-1)%\nonumber\\&+&
+e^{-i\beta}\sin \theta \,\psi_{-}(n-1,t-1)\big],
\label{Rec_P_bas}
\end{eqnarray}
and
\begin{eqnarray}
\psi^{ }_{-}(n,t)&=&e^{i\chi} \big[e^{i\beta}\sin \theta \,\psi_{+}(n+1,t-1)%\nonumber\\&-&
-e^{-i\alpha}\cos \theta\,\psi_{-}(n+1,t-1)\big],
\label{Rec_M_bas}
\end{eqnarray}
whose general solution~\cite{MM15} can be written in a compact way by using $\psi_{+}(0,0)$ and $\psi_{-}(0,0)$,
\begin{eqnarray*}
\psi_{+}(0,0)&=&\cos \eta,\\
\psi_{-}(0,0)&=&e^{i\gamma}\sin \eta,
\end{eqnarray*}
and the non-zero components of the wave function at time $t=1$,  
\begin{eqnarray*}
\psi_{+}(+1,1)&=&e^{i \chi }\left[e^{i\alpha}\cos\eta\cos \theta+ e^{i(\gamma-\beta)}\sin \eta\sin \theta\right],\\
\psi_{-}(-1,1)&=&e^{i \chi }\left[e^{i \beta}\cos\eta\sin \theta-  e^{i(\gamma-\alpha)}\sin \eta\cos \theta\right],
\end{eqnarray*}
since $\psi_+(-1,1)=\psi_-(+1,1)=0$, cf. Eqs.~\eqref{Rec_P_bas} and~\eqref{Rec_M_bas}. In terms of the preceding quantities, and for $n\in\{-t,-t+2,\cdots,t-2,t\}$, one has
\begin{equation}
\psi_{+}(n,t)=e^{i(\chi\cdot t +\alpha\cdot n)}\left[\psi_{+}(0,0)  \Lambda(n,t)+e^{-i(\chi +\alpha)}\psi_{+}(+1,1) \Lambda(n-1,t+1)\right],
\label{Sol_Psi_P}
\end{equation}
and
\begin{equation}
\psi_{-}(n,t)=e^{i(\chi \cdot t -\alpha \cdot n)}\left[\psi_{-}(0,0) \Lambda(n,t)+e^{-i(\chi -\alpha)} \psi_{-}(-1,1) \Lambda(n+1,t+1)\right],
\label{Sol_Psi_M}
\end{equation}
where
\begin{eqnarray}
\Lambda(n,t)&\equiv&\frac{1}{t+1}\Bigg\{\frac{1+(-1)^t}{2}%\nonumber \\&+&
+\sum_{r=1}^{t}  \frac{1}{\cos\omega_{r,t}}\cos\left[(t-1)\cdot \omega_{r,t}-\frac{\pi r n}{t+1}\right]\Bigg\},%\nonumber \\
\label{Lambda_def}
\end{eqnarray}
and 
\begin{equation}
\omega_{r,t} \equiv \arcsin\left(\cos\theta \sin  \frac{\pi r}{t+1}\right).
\label{omega_r}
\end{equation}
Note that in this picture the evolution of each component depends only on their own initial values. In fact, it can be shown~\cite{MM15} that $|\psi_{+}(+1,1)|^2$ can be understood as the ``rightward initial velocity'' of our quantum walker, whereas $|\psi_{-}(-1,1)|^2$ would play the role of the ``leftward initial velocity."

Even though the expression for $\Lambda(n,t)$ is completely explicit, Eq.~\eqref{Lambda_def}, it may be instructive to show how the set of equations that cross-correlate the evolution of the two components of the wave function, Eqs.~\eqref{Rec_P_bas} and~\eqref{Rec_M_bas}, turns now into a single, two-step recursive formula that governs the whole dynamics:
\begin{eqnarray}
\Lambda(n,t)&=&\cos \theta \left[\Lambda(n-1,t-1)-\Lambda(n+1,t-1)\right]+\Lambda(n,t-2).
\label{Lambda_recursive}
\end{eqnarray}
Equation~\eqref{Lambda_def} is recovered from the above relationship once one considers the initial condition $\Lambda(0,0)=1$, together with the boundary conditions $\Lambda(-n,t)=\Lambda(n,t)=0$, for $n\geq t \geq 1$.  

Observe how  $\Lambda(n,t)$ does not depend on $\chi$, $\alpha$, $\beta$, $\gamma$ or $\eta$. It is a function of $\theta$ through the value of $\cos\theta$, a property that can be also observed in Eq.~\eqref{Lambda_recursive}. One could infer from this feature that $\cos^2\theta$ plays in quantum walks the same role of $p$ in random walks, and that the rest of parameters represent mathematical degrees of freedom without correspondence in the physical world. This impression can be strengthened by computing the value of the PMF in simple examples as, for instance, when $n$ coincides with $t$: in this case $\rho_{\rm clas.}(t,t)=p^t$ while $\rho(t,t)=\cos^{2 t}\theta$.

\begin{figure}[htbp]
\begin{tabular}{rc} (a)&\includegraphics[width=0.9\columnwidth,keepaspectratio=true]{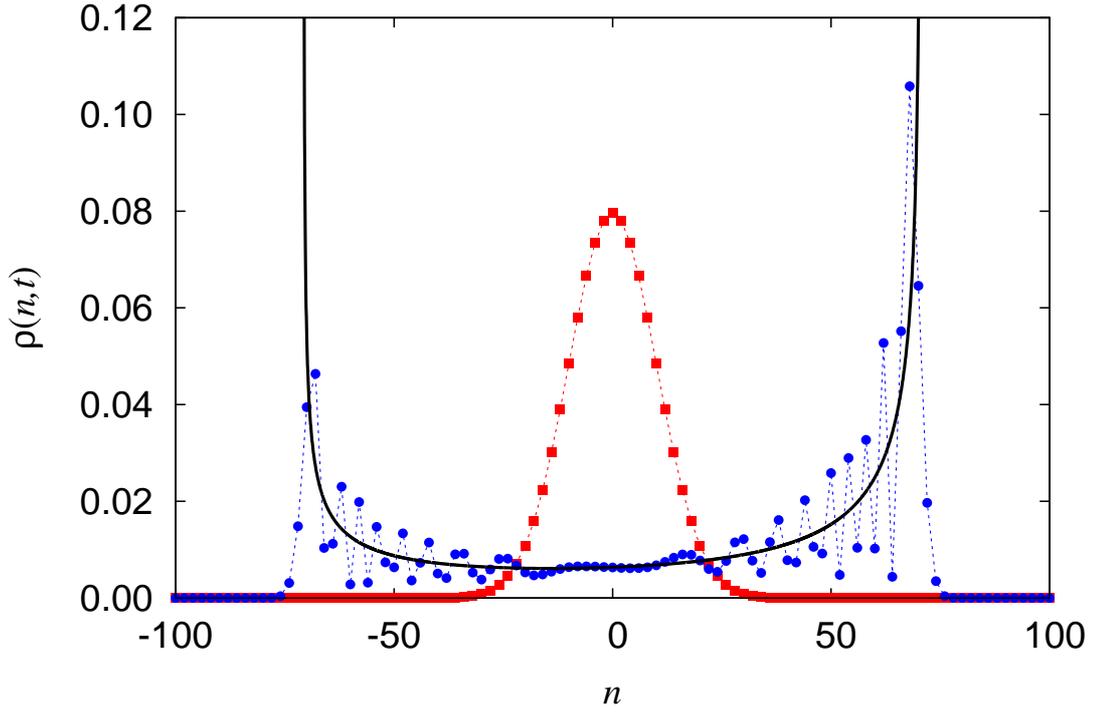}\\(b)&\includegraphics[width=0.9\columnwidth,keepaspectratio=true]{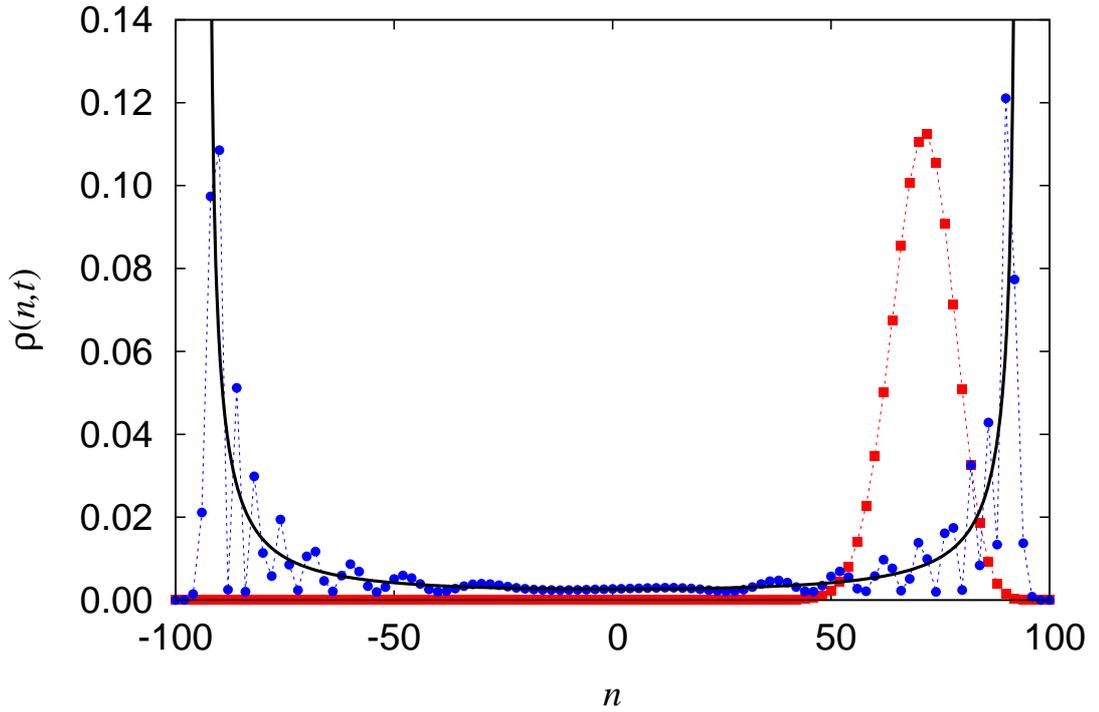}\end{tabular}
\caption{
Probability mass function after $t=100$ time steps. The dots correspond to the exact result for: (a) $\theta=\pi/4$, $\eta=\pi/16$, $\varphi=\pi$; (b) $\theta=\pi/8$, $\eta=3\pi/16$, $\varphi=\pi$; the boxes represent classical probabilities with $p=\cos^2\theta$, whereas the black solid lines correspond to $\bar{\rho}(n,t)$, cf. Eq.~\eqref{st_dens}. We have only depicted probabilities for even values of $n$, since in this case probabilities for odd values of $n$ are identically null.} 
\label{Fig_1}
\end{figure}

This conclusion is illusory, however. It is well known~\cite{TFMK03} that $\rho(n,t)$ does not depend on $\chi$, and that $\alpha$, $\beta$ and $\gamma$ appear in the PMF only in the following combination $\varphi=\alpha+\beta-\gamma$. But it is true as well that one needs to specify $\theta$, $\varphi$ and $\eta$ to determine even the most basic aspects of the evolution of quantum walks. Figure~\ref{Fig_1} illustrates this fact. In the upper panel we observe how the probability is distributed unevenly for positive and negative values of $n$, although $\theta=\pi/4$. In the lower panel we face the reversed situation, $\theta=\pi/8$ but $\rho(n,t)$ shows no clear asymmetry.

\subsection{Stationary PMF}

Figure~\ref{Fig_1} also shows us that the disparity in the bias is not the most striking aspect that distinguishes quantum walks from their classical analogues. These differences can be appreciated more easily when one considers the stationary limit~\cite{AVWW11}. It can be shown~\cite{MM15} that for $t\gg1$, the probability mass function $\rho(n,t)$ is well described by $\bar{\rho}(n,t)$,
\begin{eqnarray}
\bar{\rho}(n,t)&=&\frac{\sin \theta}{\pi} \frac{1}{t^2-n^2}\frac{1}{\sqrt{t^2\cos^2 \theta-n^2}}%\nonumber \\&\times&
\Big[t+n\left(\cos 2\eta +\sin 2\eta \tan\theta \cos\varphi\right)\Big],%\nonumber \\
\label{st_dens}
\end{eqnarray}
in the range $-t\cos\theta<n<t\cos\theta$, $0<\theta<\pi/2$. As it can be seen in Figure~\ref{Fig_1}, the agreement between $\rho(n,t)$ and $\bar{\rho}(n,t)$ is greater for small values of $n$, whereas when $|n|$ approaches to $t\cos\theta$, $\rho(n,t)$ displays an oscillatory behavior around $\bar{\rho}(n,t)$. Regardless of this, Eq.~\eqref{st_dens} captures the essence of $\rho(n,t)$:  its U-shaped profile, with a central flat region and two local maxima in the vicinity of $\pm t\cos\theta$. These traits are in clear contrast to the bell-shaped contour of the classical PMF, centered  around $\left(2 p-1\right)\cdot t$, the mean value of the displacement of the random walker, see Figure~\ref{Fig_1}. 

\begin{figure}[htbp]
\includegraphics[width=0.9\columnwidth,keepaspectratio=true]{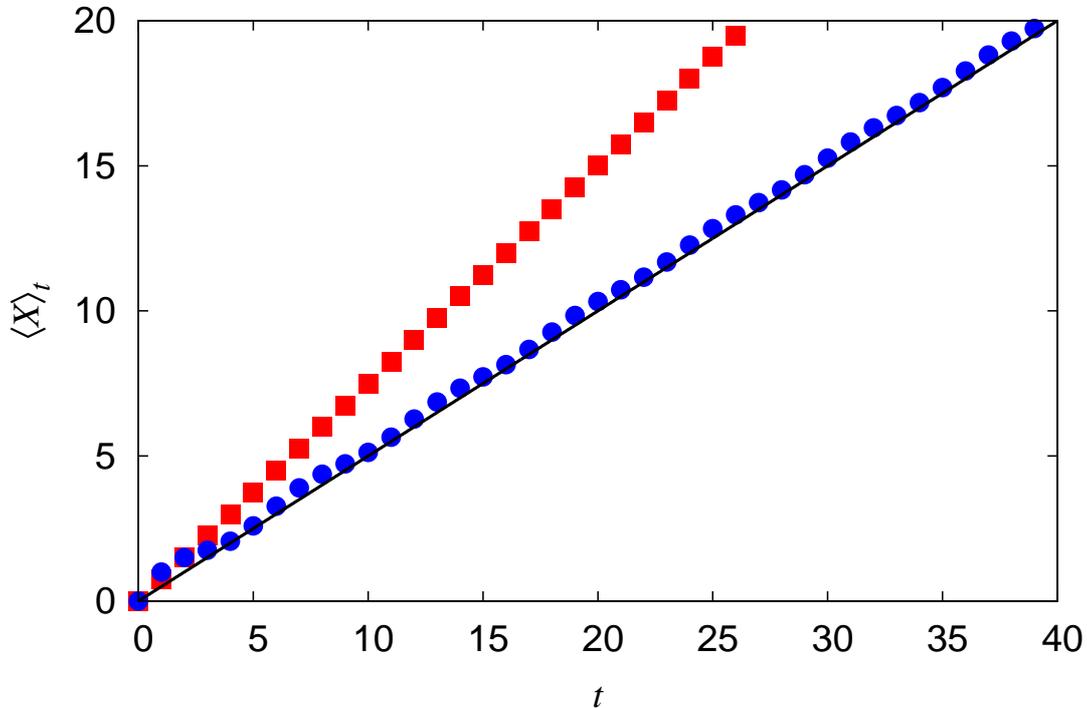}
\caption{
Expectation value of the position of the walker after $t=40$ time steps. The dots correspond to the exact result for the quantum walk with $\theta=\pi/6$, $\eta=\pi/6$, $\varphi=0$, the boxes represent the classical mean position when $p=\cos^2\theta=3/4$, whereas the black solid line corresponds to the approximate law, Eq.~\eqref{ballistic}, which in this case reads $\langle X \rangle_t  \sim  t/2$ when $\ell=1$.} 
\label{Fig_2}
\end{figure}

%\subsection{Ballistic behavior}
Regarding the expectation value of the position of the quantum walker, $\langle X \rangle_t$,
\begin{equation}
\langle X \rangle_t \equiv \ell \cdot \sum_{n=-\infty}^{\infty} n\, \rho(n,t),
\label{mean_pos}
\end{equation}
its magnitude does not stem from the location of the largest maximum of $\rho(n,t)$, but has its origin in the skewness of the distribution. An elementary analysis of $\bar{\rho}(n,t)$ reveals that any bias in $\langle X \rangle_t$ is determined in the long run by the sign of the expression between parentheses in the right hand side of Eq.~\eqref{st_dens}. Therefore, as long as
\begin{equation*}
\cos 2\eta +\sin 2 \eta\tan \theta\cos\varphi\neq0,
\end{equation*}
the expectation value of the position of the walker will increase linearly with time:
\begin{equation}
\langle X \rangle_t\sim \ell\left(1-\sin\theta\right) \left(\cos 2\eta +\sin 2\eta \tan\theta \cos\varphi\right)\, t,
\label{ballistic}
\end{equation}
as it can be checked in Figure~\ref{Fig_2}. The converse is not true~\cite{JKA12,TK12}: in order to get quantum walkers that show an exact symmetry in the parity one has to demand that  
\begin{eqnarray}
\cos 2\eta +\sin 2 \eta\tan \theta\cos\varphi=0,
\label{symm_a}
\end{eqnarray}
but also that~\footnote{Eq.~\eqref{symm_b} implies $\left|\psi_{+}(+1,1)\right|^2=\left|\psi_{-}(-1,1)\right|^2=1/2$, see Eqs.~\eqref{vel_P} and~\eqref{vel_M} below. In other words, this is the condition that ensures the absence of bias in the ``initial velocities."}
\begin{equation}
\cos 2\eta +\sin 2 \eta\tan 2\theta\cos\varphi=0,
\label{symm_b}
\end{equation}
equations that have only three main families of solutions~\cite{MM15}, being the most relevant of them the one corresponding to $\eta=\pi/4$, $\varphi=\pi/2$, for any choice of $\theta$.

\section{I\lowercase{nhomogeneous}  QW\lowercase{s}}
\label{Sec_Inhomogeneous}

The fact that not only $\chi$ but even $\alpha$ and $\beta$ (after a suitable choice of $\gamma$) can be completely ignored in the previous analysis can lead to the false conclusion that these phases can be disregarded in any other situation. We will devote the rest of this Chapter to the analysis of a framework where these magnitudes play a crucial role.

 Consider a general inhomogeneous, time-dependent unitary operator $\widehat{\mathcal{U}}_t$:
\begin{eqnarray}
\widehat{\mathcal{U}}_t
&\equiv& \sum_{n=-\infty}^{\infty}e^{i\chi^{ }_{n,t}}\big[e^{i \alpha^{ }_{n,t}} \cos \theta^{ }_{n,t} |+\rangle  \langle +| +e^{-i \beta^{ }_{n,t}} \sin \theta^{ }_{n,t} |+\rangle  \langle -| \nonumber \\
&+& e^{i \beta^{ }_{n,t}} \sin \theta^{ }_{n,t} |-\rangle  \langle +| - e^{-i \alpha^{ }_{n,t}} \cos \theta^{ }_{n,t}  |-\rangle  \langle -|\big]\otimes |n\rangle  \langle n|,%\nonumber \\
\label{U_coin_gen}
\end{eqnarray}
where $\alpha^{ }_{n,t}$, $\beta^{ }_{n,t}$, $\chi^{ }_{n,t}$ and $\theta^{ }_{n,t}$ are two-dimensional sets of real quantities. Now, we can define a new evolution operator $\widehat{\mathcal{T}}_t$, based on $\widehat{\mathcal{U}}_t$ and the standard shift operator $\widehat{\mathcal{S}}$, Eq.~\eqref{shift_operator}, $\widehat{\mathcal{T}}_t\equiv \widehat{\mathcal{S}}\, \widehat{\mathcal{U}}_t$, in such a way that the state of the particle at time $t+1$ is the result of the application of $\widehat{\mathcal{T}}_t$ to $|\psi\rangle_{t}$,  
\begin{equation}
|\psi\rangle_{t+1} =\widehat{\mathcal{T}}_t|\psi\rangle_{t}.
\label{evol_t_gen}
\end{equation}
In this case, the information supplied by the initial state of the system is not so important: Assume that $|\psi\rangle_{0}$ is of the form depicted in Eq.~\eqref{psi_zero_gen}. Then, one has that
\begin{eqnarray}
\left|\psi_{+}(+1,1)\right|^2&=&\frac{1}{2}\left[1+\cos 2\eta \cos 2 \theta^{ }_{0,0} + \sin 2\eta \sin 2 \theta^{ }_{0,0} \cos \varphi^{ }_{0,0}\right],\label{vel_P}\\
\left|\psi_{-}(-1,1)\right|^2&=&\frac{1}{2}\left[1-\cos 2\eta \cos 2 \theta^{ }_{0,0} - \sin 2\eta \sin 2 \theta^{ }_{0,0} \cos \varphi^{ }_{0,0}\right],\label{vel_M}
\end{eqnarray}
 with $\varphi^{ }_{0,0}\equiv\alpha^{ }_{0,0}+\beta^{ }_{0,0}-\gamma$. Note how this expression is invariant under the interchange
 \begin{eqnarray*}
\eta&\leftrightarrow&\theta^{ }_{0,0},\\
\gamma&\leftrightarrow&\alpha^{ }_{0,0}+\beta^{ }_{0,0}.
\end{eqnarray*}
In practice, this means that we can modify $\theta^{ }_{0,0}$ and $\alpha^{ }_{0,0}+\beta^{ }_{0,0}$ in order to obtain any desired value for $\left|\psi_{+}(+1,1)\right|$ and $\left|\psi_{-}(-1,1)\right|$, irrespective of $\eta$ and $\gamma$. The complex arguments of $\psi_{+}(+1,1)$ and $\psi_{-}(-1,1)$ can be recovered with a suitable choice of $\chi^{ }_{0,0}$ and $\alpha^{ }_{0,0}-\beta^{ }_{0,0}$.

The recursive equations of the wave-function components under the present dynamics induced by $\widehat{\mathcal{T}}_t$ are straightforward variations of Eqs.~\eqref{Rec_P_bas} and~\eqref{Rec_M_bas}:
\begin{eqnarray}
\psi^{ }_{+}(n,t)&=&e^{i\chi^{ }_{n-1,t-1}} \big[e^{i\alpha^{ }_{n-1,t-1}}\cos \theta^{ }_{n-1,t-1} \,\psi_{+}(n-1,t-1)\nonumber\\
&+&e^{-i\beta^{ }_{n-1,t-1}}\sin \theta^{ }_{n-1,t-1} \,\psi_{-}(n-1,t-1)\big],
\label{Rec_P}
\end{eqnarray}
and
\begin{eqnarray}
\psi^{ }_{-}(n,t)&=&e^{i\chi^{ }_{n+1,t-1}} \big[e^{i\beta^{ }_{n+1,t-1}}\sin \theta^{ }_{n+1,t-1} \,\psi_{+}(n+1,t-1)\nonumber\\
&-&e^{-i\alpha^{ }_{n+1,t-1}}\cos \theta^{ }_{n+1,t-1} \,\psi_{-}(n+1,t-1)\big].
\label{Rec_M}
\end{eqnarray}
Since we have a specific interest in revealing a new kind of invariance, we will introduce $\psi^{\circ}_{\pm}(n,t)$, the solution to a certain inhomogeneous, time-dependent appealing problem
\begin{eqnarray}
\psi^{\circ}_{+}(n,t)&=&e^{i\chi^{\circ}_{n-1,t-1}} \big[e^{i\alpha^{\circ}_{n-1,t-1}}\cos \theta^{ }_{n-1,t-1} \,\psi^{\circ}_{+}(n-1,t-1)\nonumber\\
&+&e^{-i\beta^{\circ}_{n-1,t-1}}\sin \theta^{ }_{n-1,t-1} \,\psi^{\circ}_{-}(n-1,t-1)\big],
\label{Rec_P0}
\end{eqnarray}
and
\begin{eqnarray}
\psi^{\circ}_{-}(n,t)&=&e^{i\chi^{\circ}_{n+1,t-1}} \big[e^{i\beta^{\circ}_{n+1,t-1}}\sin \theta^{ }_{n+1,t-1} \,\psi^{\circ}_{+}(n+1,t-1)\nonumber\\
&-&e^{-i\alpha^{\circ}_{n+1,t-1}}\cos \theta^{ }_{n+1,t-1} \,\psi^{\circ}_{-}(n+1,t-1)\big].
\label{Rec_M0}
\end{eqnarray}
Therefore, our task is to find out non-trivial relationships connecting both set of parameters. Regarding this, note that $\theta^{ }_{n,t}$ are the same in both cases: as we have seen in Section~\ref{Sec_Process}, there are some features of the process that are exclusively encoded in these magnitudes, and therefore we will exclude them from the present analysis.  

\section{I\lowercase{nvariance}}
\label{Sec_Invariance}

The properties of the system enumerated up to this point are based on the moduli of the components of the wave function. This means, in particular, that if one has that $\psi^{ }_{\pm}(n,t)$ and $\psi^{\circ}_{\pm}(n,t)$ are linked through the following identities:  %the solution to our problem reads
\begin{equation}
\psi^{ }_{+}(n,t)=\psi^{\circ}_{+}(n,t)e^{i\xi^{ }_{n,t}},
\label{psi_psi0_P}
\end{equation}
and
\begin{equation}
\psi^{ }_{-}(n,t)=\psi^{\circ}_{-}(n,t)e^{i\zeta^{ }_{n,t}},
\label{psi_psi0_M}
\end{equation}
$\rho(n,t)$ or $M(n,t)$ will remain unchanged. The new magnitudes introduced in Eqs.~\eqref{psi_psi0_P} and~\eqref{psi_psi0_M}, $\xi^{ }_{n,t}$ and $\zeta^{ }_{n,t}$, are two additional sets of arbitrary real constants, whose meaning will be discussed below.

If we assume the validity of Eqs.~\eqref{psi_psi0_P} and~\eqref{psi_psi0_M}, and replace these expressions in Eqs.~\eqref{Rec_P} and~\eqref{Rec_M}, the conditions to recover Eqs.~\eqref{Rec_P0} and~\eqref{Rec_M0} are
\begin{eqnarray*}
\chi^{\circ }_{n,t}+\alpha^{\circ }_{n,t}&=&\chi^{ }_{n,t}+\alpha^{ }_{n,t}+\xi^{ }_{n,t}-\xi^{ }_{n+1,t+1},\\
\chi^{ \circ}_{n,t}-\alpha^{\circ }_{n,t}&=&\chi^{ }_{n,t}-\alpha^{ }_{n,t}+\zeta^{ }_{n,t}-\zeta^{ }_{n-1,t+1},\\
\chi^{\circ }_{n,t}+\beta^{\circ}_{n,t}&=&\chi^{ }_{n,t}+\beta^{ }_{n,t}+\xi^{ }_{n,t}-\zeta^{ }_{n-1,t+1},\\
\chi^{\circ }_{n,t}-\beta^{\circ }_{n,t}&=&\chi^{ }_{n,t}-\beta^{ }_{n,t}+\zeta^{ }_{n,t}-\xi^{ }_{n+1,t+1}.
\end{eqnarray*}
These equations lead to the following prescription to modify the phases leaving invariant the moduli of the components of the wave function:
\begin{eqnarray}
\chi_{n,t}&=&\chi^{\circ}_{n,t}+\frac{\xi^{ }_{n+1,t+1}-\xi^{ }_{n,t}+\zeta^{ }_{n-1,t+1}-\zeta^{ }_{n,t}}{2},\label{chi_chi0}\\
\alpha_{n,t}&=&\alpha^{\circ}_{n,t}+\frac{\xi^{ }_{n+1,t+1}-\xi^{ }_{n,t}-\zeta^{ }_{n-1,t+1}+\zeta^{ }_{n,t}}{2},\label{alpha_alpha0}\\
\beta_{n,t}&=&\beta^{\circ}_{n,t}+\frac{\zeta^{ }_{n-1,t+1}+\zeta^{ }_{n,t}-\xi^{ }_{n+1,t+1}-\xi^{ }_{n,t}}{2}.\label{beta_beta0}
\end{eqnarray}

\subsection{Invariance of global observables}
\label{Sec_Quasi}

The first conclusion that can be drawn from Eqs.~\eqref{chi_chi0}--\eqref{beta_beta0} is that there is an infinite variety of choices for $\xi^{ }_{n,t}$ and $\zeta^{ }_{n,t}$ that does not modify the main properties of the quantum walker. The hard task is to identify those with a clear physical meaning or relevance. In a previous work~\cite{MM14} it has been considered one example that belongs to the following category:
\begin{eqnarray}
\xi^{ }_{n+1,t+1}&=&\xi^{ }_{n,t}, \label{prev_xi}\\
\zeta^{ }_{n-1,t+1}&=&\zeta^{ }_{n,t}. \label{prev_zeta}
\end{eqnarray}
This assumption simplifies enormously Eqs.~\eqref{chi_chi0}--\eqref{beta_beta0}:
\begin{eqnarray}
\chi^{ }_{n,t}&=&\chi^{\circ}_{n,t},\\
\alpha^{ }_{n,t}&=&\alpha^{\circ}_{n,t},\\
\beta^{ }_{n,t}&=&\beta^{\circ}_{n,t}+\zeta^{ }_{n,t}-\xi^{ }_{n,t}.\label{changing_beta}
\end{eqnarray}
One particular choice that satisfies the above requirements is $\beta^{\circ}_{n,t}=\beta^{ }_{0}$, a constant value for all $n$ and $t$, and the following functional forms for $\xi^{ }_{n,t}$ and $\zeta^{ }_{n,t}$:
\begin{eqnarray}
\xi^{ }_{n,t}&=&\frac{n-t}{2}\left(\beta^{ }_1-\beta^{ }_0\right), \label{xi_beta}\\
\zeta^{ }_{n,t}&=&\frac{n+t}{2}\left(\beta^{ }_1-\beta^{ }_0\right), \label{zeta_beta}
\end{eqnarray} 
a possible solution of Eqs.~\eqref{prev_xi} and~\eqref{prev_zeta}. The above expressions lead to the following homogeneous update rule for $\beta^{ }_{n,t}$, $t\geq 0$,
\begin{equation}
\beta^{ }_{n,t}=\beta^{ }_{t}=\beta^{ }_{0}+\left(\beta^{ }_1-\beta^{ }_0\right) t,
\label{beta_explicit}
\end{equation}
where $\beta^{ }_1$ is an arbitrary constant, whose value cannot be assessed on the basis of the knowledge of  $\rho(t,n)$, $P_{\pm}(t)$ or $M(n,t)$: it can only be inferred from the relative phase of the spinor components.

\begin{figure}[htbp]
\begin{tabular}{rc} (a)&\includegraphics[width=0.6\columnwidth,keepaspectratio=true]{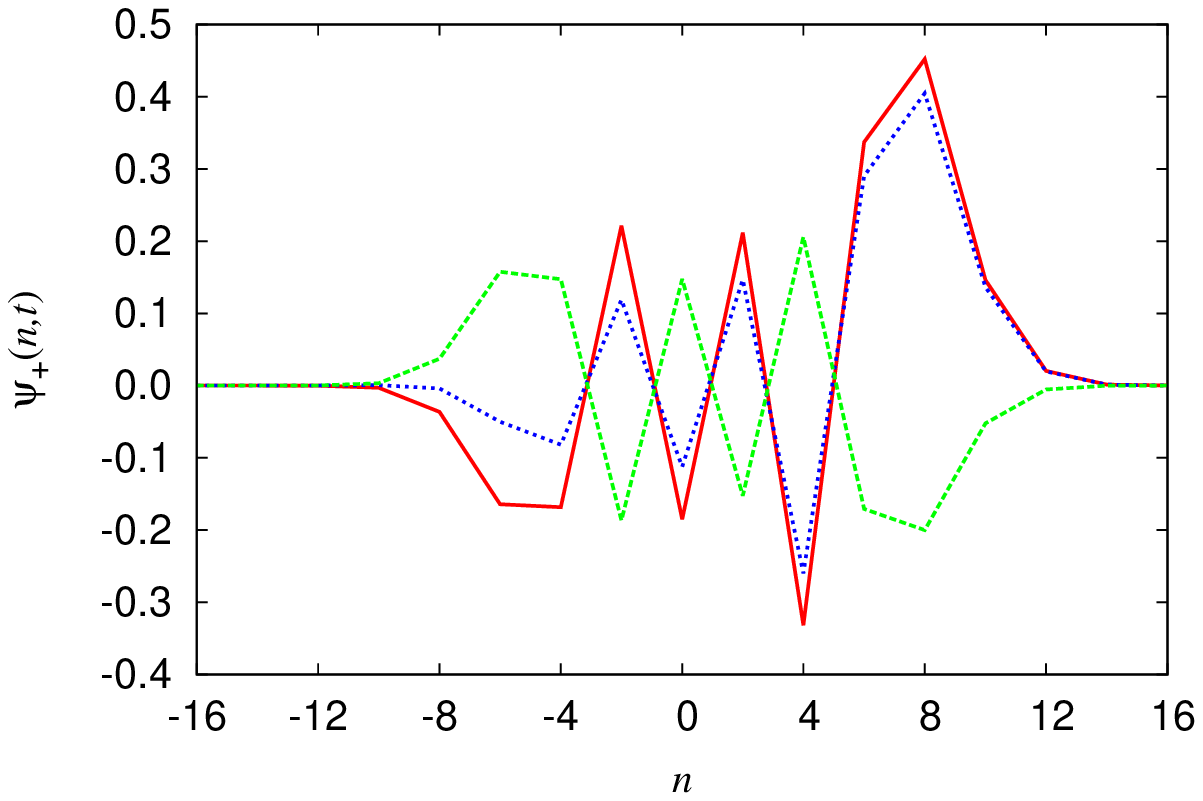}\\(b)&\includegraphics[width=0.6\columnwidth,keepaspectratio=true]{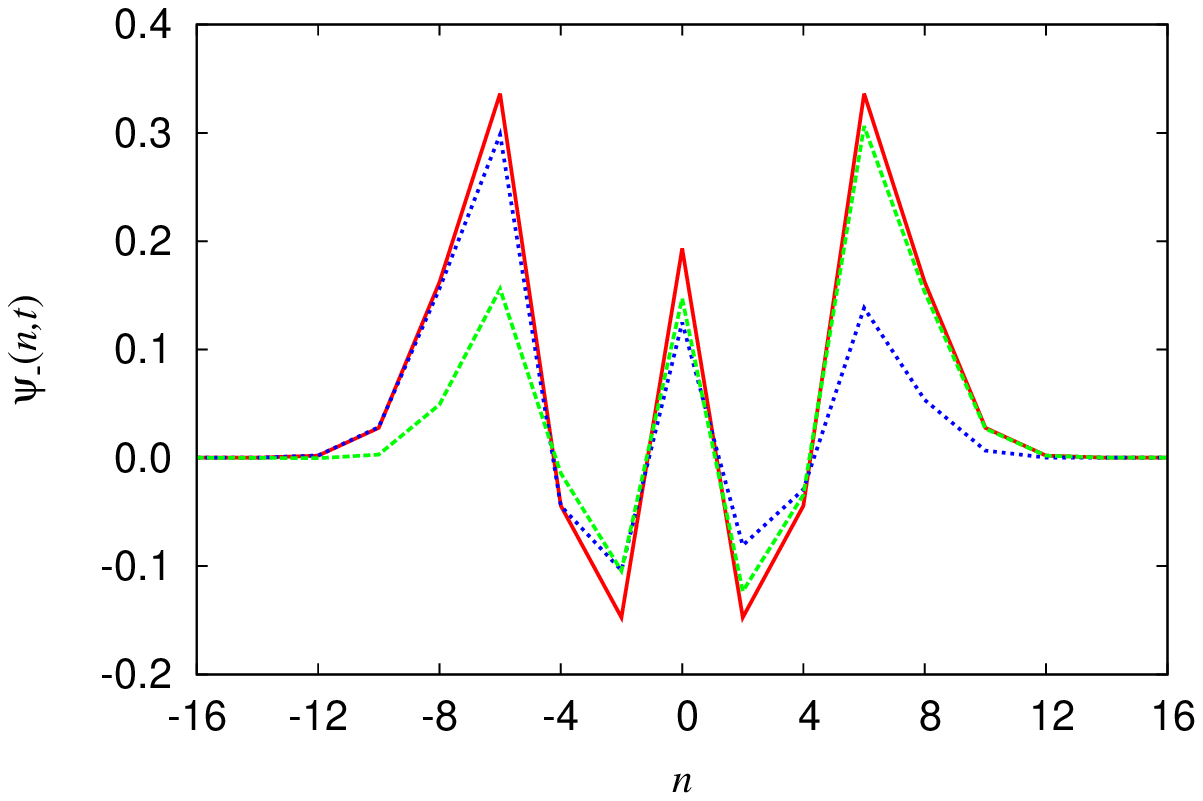}\\(c)&\includegraphics[width=0.6\columnwidth,keepaspectratio=true]{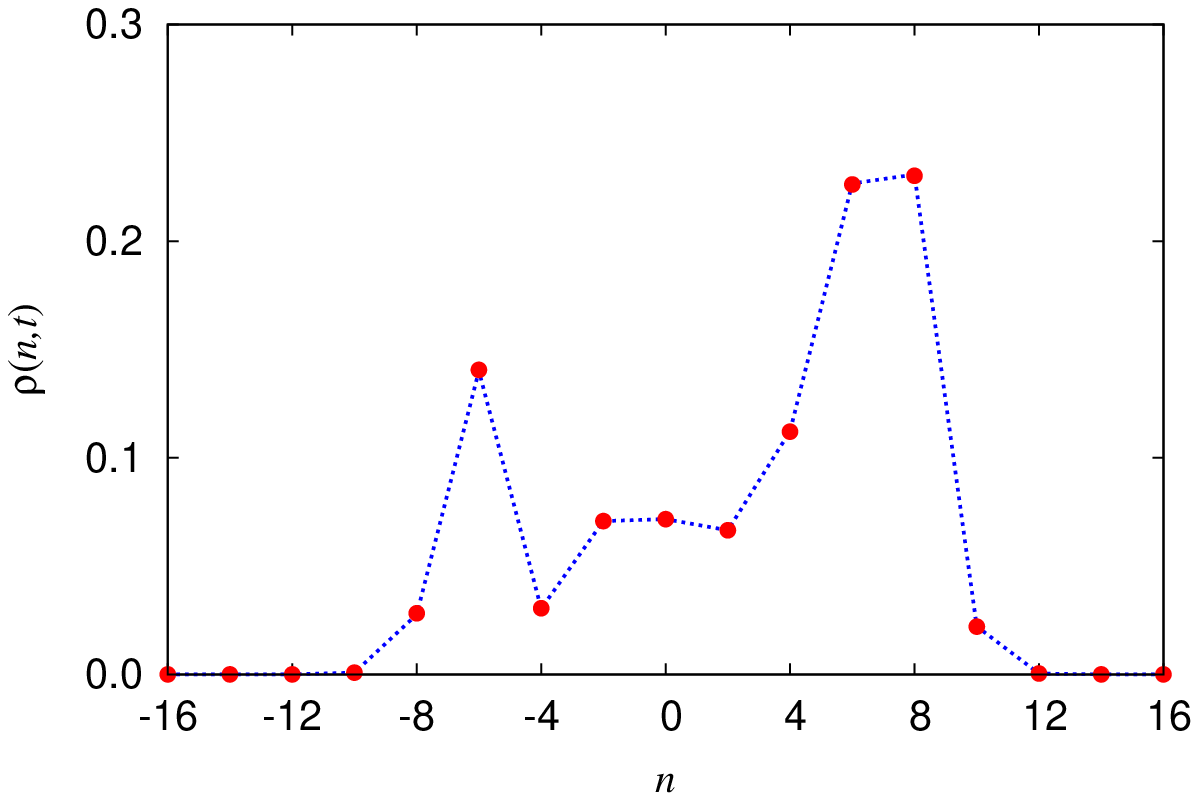} \end{tabular}
\caption{
Comparison of the wave function after $t=16$ time steps. The red solid lines and dots correspond to a time-homogeneous QW. %with $\theta=\pi/3$,  $\alpha=0$, $\beta=0$, and $\eta=\pi/3$
The blue dotted lines show the real parts of the magnitudes associated with a time-dependent QW, %with $\beta_1=1/10$, 
while the imaginary parts are depicted by green dashed lines.} 
\label{Fig_Sample}
\end{figure}

We illustrate in Figure~\ref{Fig_Sample} the invariance of $\rho(t,n)$ in spite of the time- and site-inhomogeneous phase shifts that Eq.~\eqref{beta_explicit} introduces in the wave-function components, cf. Eqs.~\eqref{xi_beta} and~\eqref{zeta_beta}. Here we have set $\theta=\pi/3$,  $\eta=\pi/3$, $\gamma=0$, $\chi=0$, $\alpha=0$, $\beta_0=0$ and $\beta_1=1/10$. With this choice, $\psi^{\circ}_{\pm}(n,t)$ are real functions that solve a stationary homogeneous problem, whereas $\psi^{}_{\pm}(n,t)$ exhibit a complex, correlated behavior: e.g., $\psi^{\circ}_{-}(n,t)$ is a symmetric function around $n=0$, while neither the real part nor the imaginary part of $\psi^{ }_{-}(n,t)$ shows this symmetry. 

We can sketch a complementary picture that may help in the understanding the behavior of $\widehat{\mathcal{U}}_t$ when $\beta_t$ follows Eq.~\eqref{beta_explicit}, through a geometrical analogy. Let us introduce $\boldsymbol{u}_t$, a time-dependent, unit-length vector in $\RR^3$. Let us denote by $\theta$ and $\beta_t$ its polar and azimuthal spherical coordinates, respectively. Then, we can recover the coin operator $\widehat{\mathcal{U}}_t$ through the scalar projection of the Pauli {\it vector\/} of operators, $\widehat{\boldsymbol{\sigma}}$, with Cartesian components %$\hat{\sigma}_j$, $j\in\{1,2,3\}$, 
\begin{eqnarray*}
\widehat{\sigma}_x&\equiv&  |+\rangle  \langle -| + |-\rangle  \langle +|,\\
\widehat{\sigma}_y&\equiv& -i |+\rangle  \langle -| + i |-\rangle  \langle +|, \mbox{ and}\\
\widehat{\sigma}_z&\equiv& |+\rangle  \langle +|  -  |-\rangle  \langle -|,
\end{eqnarray*}
onto the $\boldsymbol{u}_t$ direction,  i.e.,
\begin{equation}
\widehat{\mathcal{U}}_t\equiv \left(\boldsymbol{u}_t\cdot\hat{\boldsymbol{\sigma}}\right)\otimes\widehat{I}_P,
\end{equation}
where $\widehat{I}_P$ is the identity operator defined in the position space $\HH_P$. The evolution of $\boldsymbol{u}_t$ is a step-like precession around the north pole. Observe how, as in the example shown in Figure~\ref{Fig_Sample}, when $(\beta_1-\beta_0)/\pi$ is an irrational number, the precession of $\boldsymbol{u}_t$ is not a periodic phenomenon at all. The absence of periodicity implies that vector $\boldsymbol{u}_t$ defines an everywhere-dense but enumerable subset of points in the ring associated to colatitude $\theta$ on the sphere, and thus the unconditional probability of choosing a particular value for $\beta_t$ is uniformly distributed in the stationary limit.

\subsection{Exact invariance}
\label{Sec_Exact}
Obviously, we can go further and demand exact invariance in the problem. This can be achieved by setting $\zeta^{ }_{n,t}=\xi^{ }_{n,t}$. Eqs.~\eqref{chi_chi0}--\eqref{beta_beta0} read now~\cite{MBD14}:
\begin{eqnarray}
\chi^{ }_{n,t}&=&\chi^{\circ}_{n,t}+\frac{\xi^{ }_{n+1,t+1}+\xi^{ }_{n-1,t+1}-2\xi^{ }_{n,t}}{2},\label{chi_chi0_exact}\\
\alpha^{ }_{n,t}&=&\alpha^{\circ}_{n,t}+\frac{\xi^{ }_{n+1,t+1}-\xi^{ }_{n-1,t+1}}{2},\label{alpha_alpha0_exact}\\
\beta^{ }_{n,t}&=&\beta^{\circ}_{n,t}-\frac{\xi^{ }_{n+1,t+1}-\xi^{ }_{n-1,t+1}}{2}.\label{beta_beta0_exact}
\end{eqnarray}
As we will show below, these equations can be expressed in terms of finite differences which in turn lead to partial derivatives. In fact, in the expression of $\chi^{ }_{n,t}$ it appears a time derivative, whereas the formulas for $\alpha^{ }_{n,t}$ and $\beta^{ }_{n,t}$ contain a spacial derivative. To illustrate these statements, consider the simple choice
\begin{equation}
\xi^{ }_{n,t} = a\cdot n\cdot t. \label{a_n_t}
\end{equation}
Equations~\eqref{chi_chi0_exact} to~\eqref{beta_beta0_exact} read, as we have anticipated,
\begin{eqnarray}
\chi^{ }_{n,t}&=&\chi^{\circ}_{n,t}+a\cdot n,\\
\alpha^{ }_{n,t}&=&\alpha^{\circ}_{n,t}+a(t+1),\\
\beta^{ }_{n,t}&=&\beta^{\circ}_{n,t}-a(t+1).
\end{eqnarray}
This means, in particular, that we can transform an inhomogeneous coin into a time-dependent one
\begin{eqnarray*}
\chi^{\circ}_{n,t}=-a\cdot n &\rightarrow& \chi^{ }_{n,t}=0,\\
\alpha^{\circ}_{n,t}=0&\rightarrow& \alpha^{ }_{n,t}=a(t+1),\\
\beta^{\circ}_{n,t}=0&\rightarrow& \beta^{ }_{n,t}=-a(t+1).
\end{eqnarray*}

\subsection{Continuous limit}

Let us express Eqs.~\eqref{chi_chi0} to~\eqref{beta_beta0} in a slightly different way. Consider the discrete difference operators $\Delta_n$ and $\Delta_t$ defined as follows:
\begin{eqnarray}
\Delta_n \xi^{ }_{n,t}&\equiv&\xi^{ }_{n+1,t}-\xi^{ }_{n,t},\label{Delta_n}\\
\Delta_t \xi^{ }_{n,t}&\equiv&\xi^{ }_{n,t+1}-\xi^{ }_{n,t},\label{Delta_t}
%\Delta_n \xi^{ }_{n,t}\equiv\xi^{ }_{n+1,t}-\xi^{ }_{n,t}, &\quad &\Delta_t \xi^{ }_{n,t}\equiv\xi^{ }_{n,t+1}-\xi^{ }_{n,t},\nonumber \\
%\Delta_n \zeta^{ }_{n,t}\equiv\zeta^{ }_{n+1,t}-\zeta^{ }_{n,t}, &\quad& \Delta_t \zeta^{ }_{n,t}\equiv\zeta^{ }_{n,t+1}-\zeta^{ }_{n,t}.\nonumber
\end{eqnarray}
and similarly for $\Delta_n \zeta^{ }_{n,t}$ and $\Delta_t \zeta^{ }_{n,t}$. In terms of these operators, Eqs.~\eqref{chi_chi0} to~\eqref{beta_beta0} now read: 
\begin{eqnarray}
\chi^{ }_{n,t}&=&\chi^{\circ}_{n,t}+\frac{1}{2}\left[\Delta_n\left(\xi^{ }_{n,t+1} - \zeta^{ }_{n-1,t+1}\right)+\Delta_t \left(\xi^{ }_{n,t} + \zeta^{ }_{n,t}\right)\right],\label{chi_chi0_finite}\\
\alpha^{ }_{n,t}&=&\alpha^{\circ}_{n,t}+\frac{1}{2}\left[\Delta_n\left(\xi^{ }_{n,t+1} + \zeta^{ }_{n-1,t+1}\right)+\Delta_t \left(\xi^{ }_{n,t} - \zeta^{ }_{n,t}\right)\right],\label{alpha_alpha0_finite}\\
\beta^{ }_{n,t}&=&\beta^{\circ}_{n,t}+\zeta^{ }_{n,t}-\xi^{ }_{n,t} -\frac{1}{2}\left[\Delta_n\left(\xi^{ }_{n,t+1} + \zeta^{ }_{n-1,t+1}\right)+\Delta_t \left(\xi^{ }_{n,t} - \zeta^{ }_{n,t}\right)\right].\label{beta_beta0_finite}
\end{eqnarray}
Observe how the expression connecting $\beta^{ }_{n,t}$ and $\beta^{\circ}_{n,t}$ depends {\it explicitly\/} on $\xi^{ }_{n,t}$ and $\zeta^{ }_{n,t}$, in the sense that it is not merely a function of the increments, cf. Eq.~\eqref{changing_beta} above. In fact, we can rearrange the previous expressions in order to emphasize the distinct effects of $\xi^{ }_{n,t}$ and $\zeta^{ }_{n,t}$: 
\begin{eqnarray}
\chi^{ }_{n,t}+\alpha^{ }_{n,t}&=&\chi^{\circ}_{n,t}+\alpha^{\circ}_{n,t}+\Delta_n\xi^{ }_{n,t+1} +\Delta_t \xi^{ }_{n,t},\\
\chi^{ }_{n,t}-\alpha^{ }_{n,t}&=&\chi^{\circ}_{n,t}-\alpha^{\circ}_{n,t}-\Delta_n\zeta^{ }_{n-1,t+1}+\Delta_t  \zeta^{ }_{n,t},\\
\alpha^{ }_{n,t}+\beta^{ }_{n,t}&=&\alpha^{\circ}_{n,t}+\beta^{\circ}_{n,t}+\zeta^{ }_{n,t}-\xi^{ }_{n,t}.
\end{eqnarray}
At this point it is appropriate to note that we are not taking into account the issue of the parity of indexes $n$ and $t$: since the instances of $\xi^{ }_{n,t}$ and $\zeta^{ }_{n,t}$ that appear in Eqs.~\eqref{chi_chi0} to~\eqref{beta_beta0} are those whose subscripts have the same parity, only one of the two terms in the right-hand side of Eqs.~\eqref{Delta_n} and~\eqref{Delta_t} is relevant or even well defined. 

However, our interest in this Section is to analyze the continuous limit, $\tau\to 0$, $\ell\to 0$. Up to the first order in $\tau$ and $\ell$, one has that discrete difference operators $\Delta_n$ and $\Delta_t$ become partial derivatives:
\begin{eqnarray*}
\Delta_n &\sim&\ell \frac{\partial\mbox{\,}}{\partial X},\label{part_n}\\
\Delta_t &\sim&\tau\frac{\partial\mbox{\,}}{\partial T}.\label{part_t}
\end{eqnarray*}
We need to relate $\ell$ and $\tau$ in order to obtain an unambiguous limit. We will assume that $\ell =c\cdot\tau$, where $c$ is the characteristic speed associated with the action of the shift operator $\widehat{\mathcal{S}}$ upon the state of the walker. Therefore, depending on the physical nature of the system, $c$ represents the velocity at which the information is transferred, and it may coincide with the speed of light in vacuum. With this prescription, one has that Eqs.~\eqref{chi_chi0_finite}--\eqref{beta_beta0_finite} turn into
\begin{eqnarray}
\chi&\sim&\chi^{\circ}+\frac{\ell}{2}\left[ \partial_+ \xi+\partial_-\zeta\right],\\
\alpha&\sim&\alpha^{\circ}+\frac{\ell}{2}\left[ \partial_+ \xi-\partial_-\zeta\right],\\
\beta&\sim&\beta^{\circ}+\zeta-\xi-\frac{\ell}{2}\left[ \partial_+ \xi-\partial_-\zeta\right],
\end{eqnarray}
where $ \partial^{}_{\pm}$ are defined as follows,
\begin{equation}
\partial^{}_{\pm}\equiv\frac{\partial\mbox{\quad}}{\partial X^{\pm}}=\frac{1}{c}\cdot \frac{\partial\mbox{\,}}{\partial T}\pm \frac{\partial\mbox{\,}}{\partial X},
\end{equation}
and $X^{\pm}=c\cdot T\pm X$ are the coordinates of the null geodesics in a flat (1+1) space-time. Observe how we have removed the subscripts: the dependency on $X$ and $T$ of all the magnitudes is implicitly assumed from now on.

The exact invariance, $\zeta^{ }=\xi^{ }$, was analyzed in detail in~\cite{MBD14}. There, it is shown how the recurrence equations of the wave-function components of the walker, Eqs.~\eqref{Rec_P} and~\eqref{Rec_M}, can be mapped into equations describing the propagation of a Dirac spinor with charge $e$ and masses $m_{\pm}$ coupled to a two-dimensional Maxwell potential $\boldsymbol{A}$: 
\begin{eqnarray}
i\hbar\, \partial^{}_+ \psi^{ }_{+}+e (A_T+A_X)  \psi^{ }_{+} -  m_{+}\, c\, \psi^{ }_{-}&=&0,\label{Dirac_P}\\ 
i\hbar\, \partial^{}_- \psi^{ }_{-}+e (A_T-A_X)  \psi^{ }_{-} -  m_{-}\, c\, \psi^{ }_{+}&=&0,\label{Dirac_M}
\end{eqnarray}
whose respective space-time components must change according to the formulas
\begin{eqnarray}
A^{}_T&=&A^{\circ}_T+\frac{\hbar}{e c}\lim_{\tau\to 0} \frac{\chi-\chi^{\circ}}{\tau},\label{A_T}\\
A^{}_X&=&A^{\circ}_X+\frac{\hbar}{e c}\lim_{\tau\to 0} \frac{\alpha-\alpha^{\circ}}{\tau}.\label{A_X}
\end{eqnarray}
Note that $\zeta^{ }=\xi^{ }$ implies that
\begin{eqnarray}
\chi&\sim&\chi^{\circ}+\tau \frac{\partial \xi}{\partial {T}},\\
\alpha&\sim&\alpha^{\circ}+\ell \frac{\partial \xi}{\partial {X}},\\
\beta&\sim&\beta^{\circ}-\ell \frac{\partial \xi}{\partial {X}},
\end{eqnarray}
and, when one introduces these relationships into Eqs.~\eqref{A_T} and~\eqref{A_X} one obtains the standard gauge transformations for the components of the potential $\boldsymbol{A}$,
\begin{eqnarray}
A^{}_T&=&A^{\circ}_T+\frac{\hbar}{e c}\cdot\frac{\partial \xi}{\partial {T}},\label{gauge_AT}\\
A^{}_X&=&A^{\circ}_X+\frac{\hbar}{e}\cdot\frac{\partial \xi}{\partial {X}},\label{gauge_AX}
\end{eqnarray}
a transform that keeps invariant the electric field $E^{}_X$ acting upon the system,
\begin{equation}
E^{}_X\equiv\frac{\partial A^{}_X}{\partial {T}} -c\frac{\partial A^{}_T}{\partial {X}}=\frac{\partial A^{\circ}_X}{\partial {T}} -c\frac{\partial A^{\circ}_T}{\partial {X}}=E^{\circ}_X.
\end{equation}
If we reconsider the example introduced at the end of Section~\ref{Sec_Exact}, 
\begin{eqnarray*}
\xi&=&\frac{e E^{\circ}_X}{\hbar} \cdot X\cdot T,
\end{eqnarray*}
we can conclude that it corresponds to a case in which the electric field $E^{\circ}_X$ is constant, where we are replacing the electric potential $\phi^{\circ}$, $\phi^{\circ}=-c\cdot A^{\circ}_T$, by a time-dependent magnetic potential $A^{ }_X$,
\begin{eqnarray*}
A^{\circ}_T=-\frac{E^{\circ}_X}{c}\cdot X&\rightarrow& A^{ }_T=0,\\
A^{\circ}_X=0&\rightarrow& A^{ }_X=E^{\circ}_X \cdot T.
\end{eqnarray*}

In the most general case, when $\zeta^{ }\neq\xi^{ }$, the transformation rule for $\boldsymbol{A}$ is
\begin{eqnarray}
A^{}_T&=&A^{\circ}_T+\frac{\hbar}{2 e}\left[ \partial_+ \xi+\partial_-\zeta\right],\label{inv_AT}\\
A^{}_X&=&A^{\circ}_X+\frac{\hbar}{2 e}\left[ \partial_+ \xi-\partial_-\zeta\right],\label{inv_AX}
\end{eqnarray}
which departs from the gauge invariance of potential $\boldsymbol{A}$. However, if we investigate the change in the electric field induced by Eqs.~\eqref{inv_AT} and~\eqref{inv_AX} we find
\begin{equation}
E^{}_X-E^{\circ}_X=\frac{\partial \left[A^{}_X-A^{\circ}_X\right]}{\partial {T}} -c\frac{\partial \left[A^{}_T- A^{\circ}_T\right]}{\partial {X}}=\frac{\hbar c}{2 e} \left[ \partial_- \partial_+\xi-\partial_+\partial_-\zeta\right].\label{inv_E}
\end{equation}
Clearly,  $\zeta^{ }=\xi^{ }$ is not the only solution to the constraint
\begin{equation}
\partial_- \partial_+\xi-\partial_+\partial_-\zeta=0,\label{xi_zeta}
\end{equation}
that results in the invariance of the electric field. A possible choice is to demand that both $\xi^{ }$ and $\zeta^{ }$ satisfy the 2-dimensional wave equation by their own
\begin{equation}
\frac{1}{c^2}\frac{\partial^2 \xi}{\partial {T}^2}-\frac{\partial^2 \xi}{\partial {X}^2}=\frac{1}{c^2}\frac{\partial^2 \zeta}{\partial {T}^2}-\frac{\partial^2 \zeta}{\partial {X}^2}=0.
\end{equation}
Another alternative solution to Eq.~\eqref{xi_zeta} has appeared above, in Section~\ref{Sec_Quasi}. The equivalent expressions for Eqs.~\eqref{prev_xi} and~\eqref{prev_zeta} in the continuous limit read:
\begin{equation*}
\partial_+ \xi=\partial_-\zeta=0,
\end{equation*}
what provides another solution to Eq.~\eqref{xi_zeta}. Note that in this case Eqs.~\eqref{Dirac_P} and~\eqref{Dirac_M} show not merely covariance but perfect invariance in the mass-less case, $m_+=m_-=0$, since
\begin{eqnarray}
A^{}_T+A^{}_X&=&A^{\circ}_T+A^{\circ}_X+\frac{\hbar}{e} \partial_+ \xi,\\
A^{}_T-A^{}_X&=&A^{\circ}_T-A^{\circ}_X+\frac{\hbar}{e} \partial_-\zeta.
\end{eqnarray}

\section{C\lowercase{onclusion}}
\label{Sec_Conclusion}

Along this Chapter we have analyzed some interesting aspects of discrete-time quantum walks on the line, specifically those related with the emergence of invariance. In the first part, we have elaborated a succinct but comprehensive review covering the main features of the most elementary version of this process, when the unitary operator which assumes the function of the coin in the classical analogue is kept fixed. We have described the dynamics that determines the evolution of the walker, supplied explicit formulas for assessing the precise state of the system at any time and approximate expressions that capture the main traits of the process in the stationary limit. These equations have been very useful to pinpoint the role played by the different parameters on the solution to the problem, and put into context the generalization considered afterward. 

The second part of the Chapter contemplates the situation in which the coin is time- and site-dependent. In particular, we have focused our interest on the phase parameters that define the unitary operator and determined the constraints that must be imposed in these changing phases if one wants to obtain invariance. This invariance can be demanded at two different levels: one can require that the invariance connects states belonging to the same ray of the Hilbert space or a milder condition, that the transformation modifies unevenly the two wave-function components. In this latter case global properties (e.g., the probability that the particle is in a particular place or in a given spin state) remain unaltered but some other local quantum properties depending on the relative phase of these components can become modified. 

The Chapter ends by analyzing the introduced invariance in the continuous limit. This approach unveils that the evolution of a time- and site-inhomogeneous quantum walk can be understood in terms of the dynamics of a particle coupled to an electromagnetic field, and that the new symmetry shown by the walker can be interpreted as a manifestation of the well-known gauge invariance of electromagnetism.

%\acknowledgments
\section*{A\lowercase{cknowledgments}}
The author acknowledges partial support from the Spanish Ministerio de Econom\'{\i}a y Competitividad (MINECO) under Contract No. FIS2013-47532-C3-2-P, and from Generalitat de Catalunya, Contract No. 2014SGR608.

\end{document}